# Laser printed nano-gratings: orientation and period peculiarities

Valdemar Stankevič,[1, *] Gediminas Račiukaitis,[1] Francesca Bragheri,[2] Xuewen Wang,[3] Eugene G. Gamaly,[4] Roberto Osellame,[2] and Saulius Juodkazis[3]

[1]Center for Physical Sciences and Technology, Savanoriu Ave. 231, Vilnius LT-02300, Lithuania
[2]Istituto di Fotonica e Nanotecnologie - CNR, P.za Leonardo da Vinci 32, I-20133 Milano, Italy
[3]Center for Micro-Photonics, Faculty of Science, Engineering and Technology, Swinburne University of Technology, John St., Hawthorn, Melbourne VIC 3122, Australia
[4]Laser Physics Centre, Research School of Physics & Engineering,
The Australian National University, Canberra, Australia
(Dated: August 1, 2016)

**Understanding of material behaviour at nanoscale under intense laser excitation is becoming critical for future application of nanotechnologies. Nanograting formation by linearly polarised ultra-short laser pulses has been studied systematically in fused silica for various pulse energies at 3D laser printing/writing conditions, typically used for the industrial fabrication of optical elements. The period of the nanogratings revealed a dependence on the orientation of the scanning direction. A tilt of the nanograting wave vector at a fixed laser polarisation was also observed. The mechanism responsible for this peculiar dependency of several features of the nanogratings on the writing direction is qualitatively explained by considering the heat transport flux in the presence of a linearly polarised electric field, rather than by temporal and spatial chirp of the laser beam. The confirmed vectorial nature of the light-matter interaction opens new control of material processing with nanoscale precision.**

**Introduction**

Understanding of material behaviour at nanoscale under intense laser excitation is underpinning future laser processing technologies. Mechanical, optical, structural and compositional properties of materials could be tailored for novel alloy formation, catalytic and sensor applications. Light polarisation is an effective parameter to control the energy delivery in laser structuring of surfaces and volumes[1-5]. The orientation of self-organized deposition of materials[6], melting and oxidation of thin films by dewetting[7], laser ablation[8, 9], and self-organized ripple nano-patterns induced on the surface[10, 11] are some examples of polarisation related phenomena that gained interest recently.

The creation of surface ripples in metals or dielectric materials under laser irradiation is a well-known method to nanotexture a surface, where the ripple orientation can be finely controlled with the polarisation direction[11] and extended over two dimensions field[12, 13]. In dielectrics, nanostructuring is also possible below the surface, in the bulk of the material, by using femtosecond lasers. In particular, laser irradiation in the volume of a fused silica substrate can create self-organized nanogratings with a period in the order of a fraction of the laser wavelength[14]. Besides the fundamental interest in these nanogratings, which are the smallest structures that can be created by light in the volume of a transparent material, a few applications stemmed from these structures. In fact, it was understood that they are the basis of the microchannel formation when using the technique of femtosecond laser irradiation followed by chemical etching[15], which paved the way for the development of several optofluidic devices for biophotonic applications[16]. Another important application of nanograting formation in fused silica is the direct writing of spin-orbital polarisation converters[17], e.g. for the fabrication of $q$-plates[18]. In addition, nanograting can be exploited to write permanent optical memories with very high capacity[19]. In many of these devices, an ultrafine control of the laser-induced nanogratings is crucial. As an example, it was found that optical function of $q$-plates in silica is affected by nonhomogeneous fluorescence across the optical element due to a complex spatial pattern of the light absorbing defects[20]. This anisotropy is presumably due to a heat conduction alteration during fabrication, which affects the laser writing itself and, in the end, the performance of the optical element. Vectorial nature of light-matter interaction in the case of nanogratings[21] formation has, therefore, to be better understood.

Here, a systematic study of the nanograting width, period and orientation as a function of several irradiation parameters and most notably of the writing scan direction was carried out in fused silica, which is an isotropic matrix regarding absorption and heat diffusion. Fourier analysis of scanning electronic microscope images revealed unexpected features of the nanograting that were never reported before. While it was widely considered that nanogratings occurring perfectly perpendicular to the incident laser polarization[22, 23], however, we demonstrate that the significant tilt is observed depending on the scanning direction relative to the laser polarisation. Repeated experiments on various femtosecond laser fabrication setups and various focusing conditions were implemented, and consistently confirmed the period variations and tilting of the nanogratings for different writing directions at industrial laser printing conditions. A vectorial light-matter interaction model is put forward to explain all the observed features and to improve our understanding and control of nanograting formation.

## Results and Discussion

In the case of linear polarisation, the orientation of the nanogratings is usually predefined by the polarisation orientation, $E_y$. However, the corresponding wave vector $K$ was found to be affected by the scan orientation and was systematically studied here. Figure 1 shows a few representative examples of SEM images of the polished and wet-etched samples.
The images were used for FFT analysis to determine the tilt of the nanograting orientation $\Psi(\varphi)$ precisely for various scan directions as explained in Figure 2.

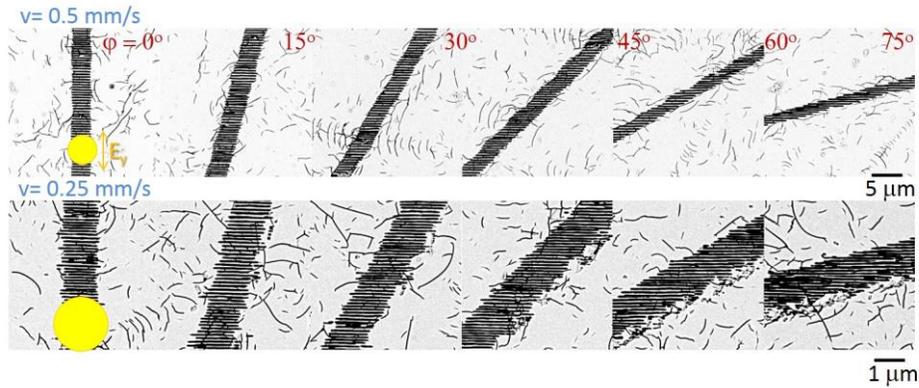

**Figure 1. SEM images of nanogratings recorded at different scan directions** $\varphi$. Processing parameters were $\lambda = 1040$ nm, $\tau_p = 317$ fs pulses (HighQ Laser) of $E_p = 600$ nJ energy (measured on target after the objective lens) at a repetition rate $f = 500$ kHz. Two different scanning speeds $v$ are reported in the two rows of images. The spot size at the focus (represented by the yellow circle in the figure) had a diameter $d = 1.22\lambda / NA \cong 2.1$ μm with $NA = 0.6$. Measurements were carried out for all 24 scan orientations (only 6 are shown here). Polarisation was fixed as $E_y$. Immersion into aqueous hydrofluoric acid solution was used to reveal the nanogratings better, but also enhanced the visibility of random scratches in the laser non-exposed surrounding areas due to non-optimal polishing process.

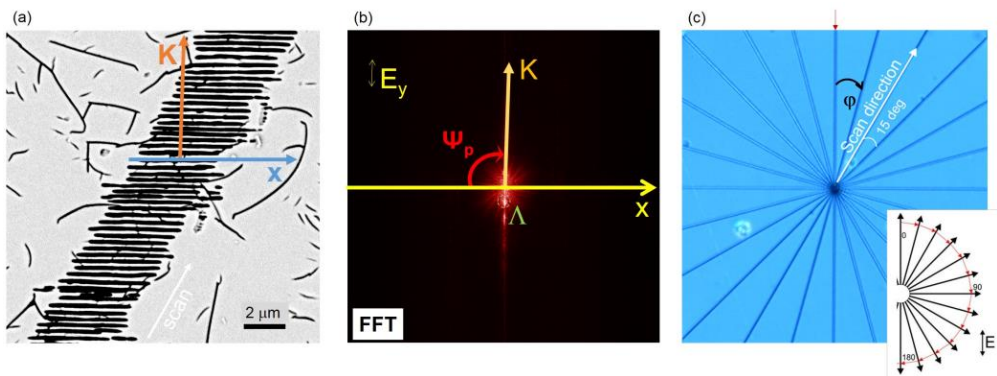

**Figure 2. Determination of the nanograting tilt angle Ψ for different scanning directions**. (a) SEM image of sub-surface nanogratings in fused silica recorded at 10 μm depth and polished afterwards for observation. The wave vector $K$ is defined as $K = 2\pi/\Lambda$, where $\Lambda$ is the nanograting period, and the $K$ direction is orthogonal to the nanograting orientation. (b) Fast Fourier transform (FFT) image of the SEM image shown in (a); polarisation $E_y$ is fixed in all experiments. The angle $\Psi_p$ is defined as the angle between the horizontal reference axis and the nanograting wave vector. (c) Optical image of the "star" pattern with the $\Delta\varphi = 15°$ angle between subsequent rays. The red arrow shows $\varphi = 0°$ position; inset shows fabrication orientation with all lines drawn from the centre outwards.

Figure 3 shows that a tilt between the nanograting orientation (wave vector) and the polarisation for different scan directions can be as high as $\Psi \sim 2°$, and the tilt angle is maximal when directions of the scan and the polarisation have an angle of $\sim \pi/4$. This tendency was observed for various scanning speeds, pulse energies, and numerical apertures at a moderate focusing.

The SEM image analysis (Fig. 1) also revealed that there was an evident difference in the width of the nanograting region depending on the scanning direction. Interestingly, FFT data showed that also the nanograting period had a remarkable angular dependence. Results of the analysis are presented in Figure. 4. A continuous change of the width of the nanostructured line, $w$, between $\varphi = 0$ and $\pi/2$ is observed, with a maximum at $\varphi = 0$. This tendency was present at different pulse energies, pulse durations (for up to twice longer pulses), focusing conditions and scanning speeds; not all results are shown for brevity. A linear dependence was observed for the modification width, $w$, on the pulse energy (Fig. 4(b)).

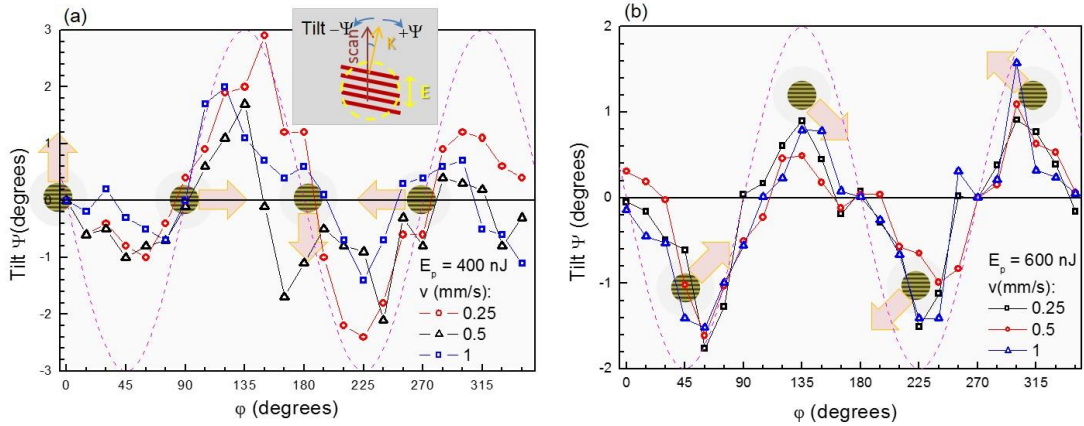

**Figure 3. Tilt angle orientational distribution $\Psi(\varphi)$ for the different scan speeds, $v$ at pulse energies.** (a) $E_p = 400$ nJ, $NA = 0.6$ and (b) $E_p = 600$ nJ, $NA = 0.4$. The inset in panel (a) defines all the relevant quantities. The dashed line is a sinusoidal function plotted as a guide for the eye. Schematic markers show the average orientation of nanogratings (corresponding to an electric field along the y-direction) while the block arrows mark the scan direction; the markers are placed at the corresponding $\varphi$ positions. This figure presents an analysis of the data partially shown in Figure 1.

A substantial change in the period of nanogratings, $\Lambda$, was observed with a strong increase at around $\varphi = \pi/2$ and $3\pi/2$ (Fig. 4(c)). At these angles, the scan direction is perpendicular to the electric field, $E_y$. On the contrary, the smallest period was observed when the scan direction was parallel to the electric field. The strong dependence of the period on the orientation of scans is intriguing since the pulse energy is maintained constant and focusing is too loose ($NA < 0.7$) to justify polarisation effects at the focal spot, as those predicted by Debye vectorial focusing[24]. The largest period $\Lambda$ occurred at the orientation of scanned lines where the width of the line, $w$, was minimal. Differences in light absorption and heat diffusion for the different scan directions have been investigated, and they are discussed in the following section.

**Theoretical model.** Formation of nanogratings inside materials[25] and on the surface[26-28] are related to the same phenomenon in the case of dielectric materials[10, 11, 29, 30]. Inside transparent materials, the period of nanogratings becomes intensity dependent via the permittivity $\varepsilon \equiv (n+ik)^2$ at the focal volume and is approximately following the $\Lambda(I) \cong (\lambda/n(I))/2$ dependence; where $n$ and $k$ are the real and imaginary parts of the refractive index, respectively.

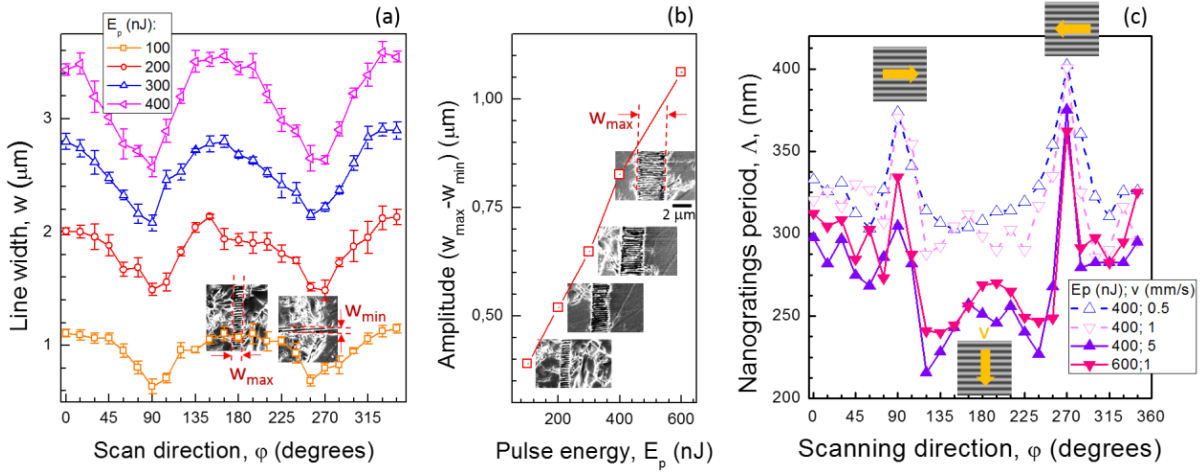

**Figure 4. General variation trend of the width and period of the nanogratings.** (a) The width of the nanogratings, $w$, vs. scanning orientation, $\varphi$, at different pulse energies, $E_p$. Focusing was $NA = 0.5$; laser pulses at $\lambda = 1030$ nm and $\tau_p = 570$ fs (Pharos laser). (b) The span of the width of nanograting line ($w_{max} - w_{min}$) at different pulse energies $E_p$. (c) Period of nanogratings, $\Lambda$, vs. scan orientation, $\varphi$, at different scan speeds, $v$. Focusing was $NA = 0.6$; laser pulses $\lambda = 1040$ nm, $\tau_p = 317$ fs (HighQ Laser). Insets in (a) and (b) show corresponding SEM images; arrow markers in (c) show the scan direction and schematic nanograting orientation. In all cases, polarisation was $E_y$.

However, a significant departure from such prediction is observed when surface plasma wave – the surface plasmon polariton (SPP) - is excited at the interface of plasma and the dielectric. The SPP wave between dielectric (glass) and plasma can be launched when $\text{Re}(\varepsilon_p) \leq -n^2$, where $\varepsilon_p$ is the permittivity of plasma at the focal volume (a necessary condition with the Bragg phase-matching being the satisfactory condition)[11, 31-34]. This follows from the requirement of the wave vector of the surface wave $k_{spp} = k_0\sqrt{\varepsilon_p\varepsilon/(\varepsilon_p+\varepsilon)}$ to be a real number, where $k_0 = 2\pi/\lambda$. Nanogratings are imprinted on the plasma-dielectric interface with a period corresponding to the half wavelength of the standing surface wave, the Bragg condition. This is why a smaller period is expected for a larger plasma density (more negative values of $(-|\text{Re}(\varepsilon_p)|)$), induced by a stronger absorption of the femtosecond laser pulses.

However, if the pulse energy is fixed, what can modulate the light absorption in the material? One simple mechanism is the different temperature of the substrate before being irradiated by the femtosecond laser pulses. In particular, if the material is hotter, the absorption is stronger and vice versa. It follows from a generic Fermi rule for optical transitions with a broader range of density of states available for the electronic transition at the higher temperature[35]. A careful analysis of the heat flow in the presence of E-field reveals that an anisotropic heat diffusion is present in an otherwise thermally isotropic material, and this causes a dependence of the absorption process on the scan direction. The heat conduction flux $q$ in a plasma placed inside an external high-frequency electric field, linearly polarised along the y-direction, has the form[36, 37]:

$$q_\alpha = -\kappa_1 \frac{\partial T}{\partial \alpha} - \kappa_2 \left| \vec{e}_\alpha \cdot \vec{e}_y \right| \frac{\partial T}{\partial y}, \quad (1)$$

where the unit vectors $\vec{e}_\alpha$ and $\vec{e}_y$ correspond to a generic direction $\alpha$ and the direction of the electric field in our case, respectively. The coefficients $\kappa_{1,2}$ are two scalar quantities obtained from the solution of the kinetic equation[37]. It follows from Eq. 1 that the heat diffusion process can be decomposed into two terms. The first one is the conventional isotropic heat diffusion while the second one is influenced by the presence of the electric field and by its direction. In particular, we can observe that the field-related term induces an enhanced heat flux in the direction of the electric field while its contribution vanishes in the direction orthogonal to the electric field. The heat affected zone, after the creation of a plasma in the focal volume is therefore asymmetrically shaped with an elliptical cross section (see Fig. 5(a)), where the major axis is aligned with the linearly polarised electric field. The manifestation of such heat-enhanced diffusion along the E-field in 3D laser printing has been confirmed in a two-photon polymerization process[5]. Here, the asymmetry of the heated zone can explain the observed variations in the nanograting period, width and tilt. These aspects are discussed in details in the following sections.

**Explanation of nanograting period and line-width dependencies.** Figure 4 shows a general trend with concerning the pulse energy used to inscribe the nanograting. In fact, for an increasing pulse energy, the width of the written lines expands and the nanograting period shrinks for all scanning directions. This general behaviour is explained by considering that the absorption process is nonlinear. Hence a stronger pulse intensity can broaden the absorption volume (thus increasing the width of the nanograting region) and also increase the plasma density and therefore reduced transmission[38] (thus reducing the nanograting period, as discussed in the previous section). These two effects are rather straightforward, but they do not account for the observation that width and period of the nanogratings can also vary, at the same pulse energy, for different scan directions. We address this aspect now providing a simple explanation based on the theoretical model previously presented. The largest periods of the nanogratings were observed at the scan orientations equal to $\varphi = \pi/2$ and $3\pi/2$, in correspondence to the minimum widths of the modified region. As previously mentioned, during the sample scanning, the laser absorption is affected by the pristine temperature of the material where the subsequent pulse impinges. However, as shown in Eq. 1, the heat diffusion process during the laser irradiation is anisotropic, and this creates a heat affected zone that is elliptical in the plane orthogonal to the direction of the laser beam propagation (see Fig. 5(a)). As a consequence, the beam moving in different scanning directions will encounter material that has been more or less pre-heated, and thus prepared to a larger or weaker absorption. Figure 5(b-d) clearly visualise three different situations. When the scanning direction is along the electric field ($\varphi = 0, \pi$), the subsequent pulses will encounter a pre-heated substrate and will experience the maximum absorption. When $\varphi = \pi/2$ and $3\pi/2$, instead, the absorption is at the minimum, while, in all other directions, we have an intermediate behaviour. Since larger absorptions cause wider modifications with smaller nanograting periods, as already discussed, this mechanism fully explains the observed dependencies on the scanning directions.

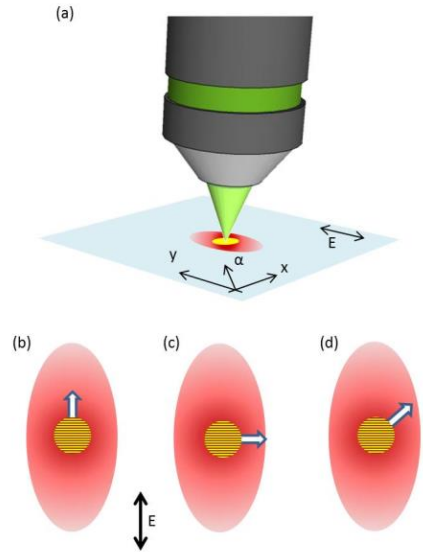

**Figure 5. Schematics of thermal diffusion process affected by the coupling between the plasma electrons and the electric field**. (a) The anisotropic heat affected zone (red region) due to an enhanced heat flux along the electric field direction. Yellow spot represents the plasma in the focal volume. (b)- (d) Three different scan directions (represented by the arrows) oriented relative to the heat affected region. The average orientation of the nanogratings is also schematically reported as a reference.

**Explanation of nanograting tilt.** The tilt of the nanograting orientation is an even more puzzling phenomenon that is observed here for the first time (Fig. 3). No tilt of the nanograting orientation is observed for scanning directions corresponding to $\varphi = 0, \pi/2, \pi$ and $3\pi/2$; while the maximum rotation is observed for $\varphi = \pi/4, 3\pi/4, 5\pi/4$ and $7\pi/4$.

Since the formation of nanogratings is defined primarily by the electronic plasma excitation, we investigated at first the possible role of the temporal and spatial chirp of the ultra-short laser pulses. Temporal and spatial chirps were measured, varied and correlated with the tilt and period of nanogratings (see Supplementary Information for details). Both the temporal and spatial chirps were found to influence the tilt of the nanogratings, but only at very large chirp values. In the conditions used for the experiments presented here, the pulse duration was the shortest, and the pulse front tilt was negligible. Therefore, temporal and spatial chirps cannot be invoked to explain the nanograting tilt dependence on the scanning direction.

Actually, the same theoretical model, used to explain the dependencies of the width and period of the nanogratings, can also explain the tilting effect. The conditions where no tilt was observed correspond to a symmetric heat affected zone with respect to the scanning direction (see Fig. 5(b, c)), while the maximum tilt was observed where the heat affected zone has the largest unbalance with respect to the scanning direction. A symmetric heat affected zone means that the subsequent pulses will hit an evenly preheated material, and thus, the absorption process will be the same in the whole focal volume, and the symmetry of the process will force the nanograting wave vector ***K*** to be parallel or orthogonal to the polarisation. On the contrary, when the heat affected zone is asymmetric with respect to the scanning direction, absorption will be different on the two sides of the focal volume, and this will induce a nanograting period that is shorter (longer) in the hotter (colder) side. As a consequence, the overall nanograting orientation will be affected, with a wave vector ***K*** rotated toward the hotter side.

As an example, let's consider Fig. 5(d), where we schematically represented the situation for $\varphi = \pi/4$. In this case, each new-coming light pulse meets the material that it hotter on the left side of the scanning direction and colder on the right side. For this reason, the periodicity of the nanogratings (generally orthogonal to the electric field direction) is slightly reduced on the left side comparing to the right one. This results in a negative tilt $\Psi < 0$ of the nanogratings for this scanning direction, which is exactly what was observed experimentally (Fig. 3).

Figure 3 also shows that the amplitude of the nanograting tilt increases with the pulse fluence (compare panel (a) and (b), where we have used a fluence of 11.4 J/cm$^2$ and 7.6 J/cm$^2$, respectively), while the dependency on the scanning direction is the same. This further feature can also be explained by the above model. In fact, a larger pulse fluence means a stronger temperature gradient induced in the focal volume. According to Eq. 1, this corresponds to a stronger contribution of the field-related term in the heat diffusion process, resulting in a more elliptical heat affected region. This causes even stronger unbalance of material temperature on the two sides of the scanning direction at $\varphi = \pi/4$, thus inducing a stronger rotation of the nanograting orientation. Consistently with the proposed model, at the larger pulse fluence, a larger tilt was observed.

**Conclusions**

A systematic study of the main nanograting features relative to the direction of the laser beam scanning was carried out on a broad range of the parameter space. In particular, different pulse energies, scanning speeds, focusing, and temporal and spatial chirps have been investigated. In all these conditions, we have shown for the first time a reproducible dependence of the nanograting width, period and tilt on the writing scan direction, which can affect the performance of directly written photonic components based on the properties of the nanogratings, as for example the laser-written *q*-plates.

For linearly polarised laser pulses, the strongest variations of the nanograting width and period were observed for the scanning directions parallel and perpendicular to the electric field direction while the maximum tilt of the nanograting orientation was observed when the scanning direction was at $\varphi = 45°$ relative to the electric field direction. All these observations can be consistently explained by an anisotropic heat-diffusion model that takes into account coupling of the hot electrons in the plasma with the pulse electric field, enhancing the heat diffusion in the direction of the latter. This anisotropic heating of the substrate is responsible for a modulated absorption of light along the different scan directions and explains all the observed features.

The experimental results here reported represent a clear evidence of a polarization-affected light-matter interaction process. The observed features are expected to be even richer in the case of vector beams and at the tight focusing, where vectorial nature of light has a strong presence. This work paves the way to a clearer understanding of these phenomena and their exploitation for optimisation of current devices and the design of innovative ones taking full advantage of the vectorial aspects of light-matter interaction.

**Methods**

**Fabrication of sub-surface nanogratings.** Two different femtosecond lasers were used to record sub-surface nanogratings: *(1)* Pharos (Light Conversion) with the wavelength of 1030 or 515 nm, and the 260 fs laser pulses at the 500 kHz repetition rate and the scanning speed of 0.25 mm/s and 1 mm/s; *(2)* FemtoRegen (HighQ Laser) 1040 nm, 317 fs at the 500 kHz repetition rate and the scanning speeds from 0.25 to 5 mm/s. Focusing was carried out in case *(1)* with a 50× objective lens of numerical aperture $NA = 0.55$ (Olympus LMPlan). For case *(2)*, the employed objectives had $NA = 0.6$ at 50× magnification (Leitz Wetzlar) or $NA = 0.4$, at 20× (Olympus LMPlan).

The pulse energy was measured after the objective lens at the sample location. Nanogratings were recorded in a multi-shot exposure regime, e.g., for a typical spot diameter at the focus of 2.5 μm, $f = 500$ kHz laser repetition rate and a typical scan speed of $v = 1$ mm/s, there were $N \cong 1.3 \times 10^3$ pulses per spot.

Nanogratings were recorded at 10 μm depth below the surface of ultraviolet-grade fused silica glass (JGS1), and samples were mechanically polished to the depth of strongest modification. A short immersion into 5%wt. aqueous solution of hydrofluoric acid was used to facilitate the surface morphology analysis by scanning electron microscopy (SEM); a 5 nm thick gold coating was used for SEM imaging.

**Estimation of the nanograting tilt angle.** In order to determine the angle $\Psi$ between the orientation of nanogratings and polarisation for different scanning directions, the following procedure was carried out (see Fig. 2). The orientation angle $\Psi_p$ of the wave vector $K = 2\pi/\Lambda$ of the nanogratings was determined for each scan direction with respect to the SEM image x-axis. There were 24 scan directions, $\varphi$, with $\Delta\varphi = 15°$ separation between the neighbouring rays (Fig. 2(c)). A fixed polarisation, $E_y$, was used to write all the lines, i.e. a polarisation parallel to the $K$ wave vector at $\varphi = 0°$. The tilt angle of the nanograting wave vector with respect to the polarization direction was calculated as $\Psi_p(0°) - \Psi_p(\varphi)$ for the various scanning directions in order to compensate for possible misalignments between the image x-axis and the $\varphi = 90°$ orientation when placing the sample in the SEM; by this definition, the positive tilt $+\Psi$ corresponds to a clockwise (cw) rotation of the nanograting orientation. To further reduce errors in positioning and judgement of nanograting orientation, the Fast Fourier Transform (FFT) of the images were calculated in random and sequential order with respect to the $\varphi$ angles. In addition, a test of two different people carrying out the analysis using the same SEM images with Gwyddion and ImageJ freeware packages was used as a reference test.

**Acknowledgements**
Authors are grateful to Linas Giniunas for discussions on spatial chirp of Pharos laser pulses and Algirdas Selskis for help in taking multiple SEM images of laser-written stars. Partial support via ARC Discovery DP120102980 is acknowledged.

**Author Contributions**
V.S. and G.R. observed for the first time the nanograting tilt. V.S., after discussion with S.J., designed the experiment, fabricated the samples, measured and analysed all experimental data. R.O. together with F.B. provided the laboratory for experiments with a different setup. F.B. partly participated in the sample fabrication and data analysis. S.J. provided the initial phenomena explanation ideas, which were further discussed and refined together with R.O. X.W. performed the initial numerical simulations for phenomena understanding. E.G.G. provided advice and helpful discussion on the model. All authors discussed the results and contributed to the writing of the paper.

**Competing financial interests:** The authors declare no competing financial interests.